\documentclass[letterpaper,table]{zadpreprints}
\pdfoutput=1

\usepackage{ragged2e}
\sectionfont{\RaggedRight}

\newcommand{\eightpt}{\fontsize{7}{5}\selectfont}
\renewcommand{\footnotesize}{\eightpt}

\makeatletter 
\renewcommand{\@makefntext}[1]{%
  \setlength{\parindent}{0pt}%
  \begin{list}{}{\setlength{\labelwidth}{2mm}
    \setlength{\leftmargin}{40pt}%
    \setlength{\labelsep}{1pt}%
    \setlength{\itemsep}{0pt}%
    \setlength{\parsep}{0pt}%
    \setlength{\topsep}{0pt}%
    \color{black}\footnotesize}%
  \item[\@thefnmark\hfil]#1
  \end{list}%
}
\makeatother 

\graphicspath{ArXiV/}
\DeclareUnicodeCharacter{2212}{-}

\papertype{Applied Statistics}

\title{\fontsize{12}{18}\selectfont Semantic and Cognitive Tools to Aid Statistical Science:\\ \fontsize{11}{17}\selectfont Replace Confidence and Significance by Compatibility and Surprise}

\citearticle{Rafi, Z. \& Greenland, S. Semantic and Cognitive Tools to Aid Statistical Science: Replace Confidence and Significance by Compatibility and Surprise. \textit{\href{https://arxiv.org/abs/1909.08579}{arXiv:1909.08579 [stat.ME]}} (2020).}

\author[1]{Zad Rafi \protect\orcidicon{0000-0003-1545-8199}}
\author[2]{Sander Greenland \protect\orcidicon{0000-0003-4364-3279}}

\affil[1]{Department of Population Health, NYU Langone, New York, NY}
\affil[2]{Department of Epidemiology and Department of Statistics, 
University of California, Los Angeles, CA}

\corrcontact{$^{1}$Email: zad@lesslikely.com \\$^{1}$Twitter: \href{https://twitter.com/DailyZad}{@DailyZad} \\$^{2}$Email: lesdomes@ucla.edu \\$^{2}$Twitter: \href{https://twitter.com/Lester_Domes}{@Lester\_Domes}}
\abbreviation{\scriptsize {A}: Background model assumptions;\\ {CI}: Compatibility/confidence interval;\\ {H}: Test hypothesis;\\ {LI}: Likelihood interval;\\ {LR}: Likelihood ratio;\\ {M}: Test model;\\ {MLR}: Maximum-likelihood ratio;\\ {\textit{S}-value}: Surprisal value}

\hypersetup{
pdftitle={Semantic and Cognitive Tools to Aid Statistical Science: Replace Confidence and Significance by Compatibility and Surprise},
pdfsubject={Statistical Science, Applied Statistics},
pdfauthor={Zad Rafi and Sander Greenland},
pdfdate={2020-09-23},
pdfproducer=pdfLaTeX,
pdfcopyright={Copyright (C) 2020,  The Author(s), BMC Medical Research Methodology},
pdfkeywords={Interval Estimates, P-values, S-values, Information Statistics, Bayesian Statistics, Statistical Models, Data Interpretation, Statistical Significance, Cognitive Science},
pdflang={en},
pdfmetalang={en}
}

\begin{document}

\maketitle

\begin{abstract}\vspace{-0.035cm}
{\setlength{\parskip}{.70em}
{\bfseries\fontsize{6}{12}\allcaps{Background}}: {\fontsize{7}{12}\selectfont{Researchers often misinterpret and misrepresent statistical outputs. This abuse has led to a large literature on modification or replacement of testing thresholds and \textit{P}-values with confidence intervals, Bayes factors, and other devices. Because the core problems appear cognitive rather than statistical, we review some simple methods to aid researchers in interpreting statistical outputs. These methods emphasize logical and information concepts over probability, and thus may be more robust to common misinterpretations than are traditional descriptions.}}\par\noindent
{\bfseries\fontsize{6}{12}\allcaps{Methods}}: {\fontsize{7}{12}\selectfont{We use the Shannon transform of the \textit{P}-value $p$, also known as the binary surprisal or \textit{S}-value $s=−\log_{2}(p)$, to provide a measure of the information supplied by the testing procedure, and to help calibrate intuitions against simple physical experiments like coin tossing. We also use tables or graphs of test statistics for alternative hypotheses, and interval estimates for different percentile levels, to thwart fallacies arising from arbitrary dichotomies. Finally, we reinterpret \textit{P}-values and interval estimates in unconditional terms, which describe compatibility of data with the entire set of analysis assumptions. We illustrate these methods with a reanalysis of data from an existing record-based cohort study.}}\par\noindent
{\bfseries\fontsize{6}{12}\allcaps{Conclusions}}: {\fontsize{7}{12}\selectfont{In line with other recent recommendations, we advise that teaching materials and research reports discuss \textit{P}-values as measures of compatibility rather than significance, compute \textit{P}-values for alternative hypotheses whenever they are computed for null hypotheses, and interpret interval estimates as showing values of high compatibility with data, rather than regions of confidence. Our recommendations emphasize cognitive devices for displaying the compatibility of the observed data with various hypotheses of interest, rather than focusing on single hypothesis tests or interval estimates. We believe these simple reforms are well worth the minor effort they require.}}
}

\vspace{0.2cm}
\end{abstract}

\begin{multicols}{2}

\section{Background}

\vspace{-.5cm}
\lettrine[lines=3]{S}{tatistical} science is fraught with psychological as well as technical difficulties, yet far less attention has been given to cognitive problems than to technical minutiae and computational devices \cite{Greenland2017-es,gigerenzerMindlessStatistics2004}. If the issues that plague science could be resolved by mechanical algorithms, statisticians and computer scientists would have disposed of them long ago. But the core problems are of human psychology and social environment, one in which researchers apply traditional frameworks based on fallacious rationales and poor understanding \cite{Greenland2017-es,Stark2018-eo}. These problems have no mathematical or philosophical solution, and instead require attention to the unglamorous task of developing tools, interpretations and terminology more resistant to misstatement and abuse than what tradition has handed down. \\ \indent
We believe that neglect of these problems is a major contributor to the current crisis of statistics in science \cite{Simmons2011-em,Open_Science_Collaboration2015-ph,Freedman2015-bz,Camerer2016-yf,Lash2018-lf,cassidyFailingGrade892019}. Several informal descriptions of statistical formulas may be reasonable when strictly adhered to, but nevertheless lead to severe misinterpretations in practice. Users tend to take extra leaps and shortcuts, hence we need to anticipate implications of terminology and interpretations to improve practice. In doing so, we find it remarkable that the \textit{P}-value is once again at the center of the controversy \cite{leekStatisticsValuesAre2015}, despite the fact that some journals strongly discouraged reporting \textit{P}-values decades ago \cite{Lang1998-sb}, and complaints about misinterpretation of statistical significance date back over \textit{a century} \cite{pearsonNoteSignificantNonsignificant1906,boringMathematicalVsScientific1919,tylerWhatStatisticalSignificance1931}. Equally remarkable is the diversity of proposed solutions, ranging from modifications of conventional fixed-cutoff testing \cite{Benjamin2017-fz,Lakens2018-lg,Lakens2018-im,mayoStatisticalInferenceSevere2018} to complete abandonment of traditional tests in favor of interval estimates \cite{rothmanShowConfidence1978,blandMeasuringAgreementMethod1999,cummingUnderstandingNewStatistics2012} or testing based on Bayesian arguments \cite{colquhounFalsePositiveRisk2019,goodmanIntroductionBayesianMethods2005,heldNewStandardAnalysis2020,matthewsMovingPost052019,sellkeCalibrationValuesTesting2001}; no consensus appears in sight.  \\ \indent
While few doubt that some sort of reform is needed, the following crucial points are often overlooked: 
\vspace{-0.25cm}
\begin{enumerate}
\item There is no universally valid way to analyze data and thus no single solution to the problems at hand. 
\item Careful integration of contextual information and technical considerations will always be essential.
\item Most researchers are under pressure to produce definitive conclusions, and so will resort to familiar automated approaches and questionable defaults \cite{wangResearcherRequestsInappropriate2018}, \textbf{with} or \textbf{without} \textit{P}-values or “statistical significance" \cite{gelmanProblemsPvaluesAre2016}. 
\item Most researchers lack the time or skills for re-education, so we need methods that are simple to acquire quickly based on what is commonly taught, yet are also less vulnerable to common misinterpretation than are traditional approaches (or at least have not yet become as widely misunderstood as those approaches).
\vspace{-0.25cm}
\end{enumerate}

Thus, rather than propose abandoning old methods in favor of entirely new ones, we will review simple cognitive devices, terminological reforms, and conceptual shifts that encourage more realistic, accurate interpretations of conventional statistical summaries. Specifically, we will advise that: 
\vspace{-0.25cm}
\begin{itemize}
\item We should replace decisive-sounding, overconfident terms like “significance,” “nonsignificance” and “confidence interval,” as well as proposed replacements like “uncertainty interval,” with more modest descriptors such as “low compatibility,” “high compatibility” and “compatibility interval”  \cite{amrheinScientistsRiseStatistical2019,greenlandValidPvaluesBehave2019,greenlandAreConfidenceIntervals2019}. 
\item We should teach alternate ways to view \textit{P}-values and interval estimates via information measures such as \textit{S}-values (surprisals), which are the negative logarithms of the \textit{P}-values; these measures facilitate translation of statistical test results into results from simple physical experiments \cite{greenlandValidPvaluesBehave2019,coleSurprise2020}. 
\item For quantities targeted for study, we should replace single \textit{P}-values, \textit{S}-values, and interval estimates by tables or graphs of \textit{P}-values or \textit{S}-values showing results for relevant alternative hypotheses as well as for null hypotheses.  
\item We should from the start teach that the usual interpretations of statistical outputs are often misleading even when they are technically accurate. This is because they \textit{condition} on background assumptions (i.e., they treat them as given), and thus they ignore what may be serious uncertainty about those assumptions. This deficiency can be most directly and nontechnically addressed by treating them \textit{unconditionally}, shifting their logical status from assumptions to components of the tested framework. \\
\vspace{-0.85cm} 
\end{itemize}

We have found that the last recommendation (to decondition inferences \cite{greenlandValidPvaluesBehave2019}) is the most difficult for most readers to comprehend, and is even resisted and misrepresented by some with extensive credentials in statistics. Thus, to keep the present paper of manageable length we have written a companion piece, \href{https://arxiv.org/abs/1909.08583}{\textbf{Greenland \& Rafi, 2020}} \cite{greenlandAidScientificInference2020}, which explains in depth the rationale for de-emphasizing traditional conditional interpretations in favor of unconditional interpretations.

\noindent\textcolor{white}{\rule[-0.5ex]{\linewidth}{0.05pt}}
\vspace{-1.25cm}
\section{An Example}
\vspace{-0.150cm}
We will display some of these problems and recommendations with published results from a record-based cohort study of serotonergic antidepressant prescriptions during pregnancy and subsequent autism spectrum disorder (ASD) of the child (Brown et al. \cite{brownAssociationSerotonergicAntidepressant2017}). Out of 2,837 pregnancies that had filled prescriptions, approximately 2\% of the children were diagnosed with ASD. The paper first reported an adjusted ratio of ASD rates (hazard ratio or HR) of 1.59 when comparing mothers with and without the prescriptions, and 95\% confidence limits (CI) of 1.17 and 2.17. This estimate was derived from a proportional-hazards model which included maternal age, parity, calendar year of delivery, neighborhood income quintile, resource use, psychotic disorder, mood disorder, anxiety disorder, alcohol or substance use disorder, use of other serotonergic medications, psychiatric hospitalization during pregnancy, and psychiatric emergency department visit during pregnancy. \\ \indent
The paper then presented an analysis with adjustment based on a high-dimensional propensity score (HDPS), in which the estimated hazard ratio became 1.61 with a 95\% CI spanning from 0.997 to 2.59. Despite the estimated \textit{61\% increase} in the hazard rate in the exposed children and an interval estimate including ratios as large as 2.59 and no lower than 0.997, the authors still declared that there was no association between in utero serotonergic antidepressant exposure and ASD because it was not “statistically significant.” This was a misinterpretation of their own results because an association was not only present, but also quite close to the 70\% increase they reported from previous studies \cite{brownAssociationAntenatalExposure2017}. Yet the media simply repeated Brown et al.’s misstatement that there was no association after adjustment \cite{yasgurAntidepressantsPregnancyNo2017}. \\ \indent
This type of misreporting remains common, despite the increasing awareness that such dichotomous thinking is detrimental to sound science and the ongoing efforts to retire statistical significance \cite{goodmanIntroductionBayesianMethods2005,amrheinScientistsRiseStatistical2019,altmanAbsenceEvidenceNot1995,amrheinEarthFlat052017,mcshaneAbandonStatisticalSignificance2019,pooleConfidenceIntervalsExclude1987,rothmanSignificanceQuesting1986,wassersteinMovingWorld052019}. To aid these efforts, we will explain the importance of showing results for a range of hypotheses, which may help readers see why conclusions such as in Brown et al. \cite{brownAssociationSerotonergicAntidepressant2017,yasgurAntidepressantsPregnancyNo2017} represent dramatic misinterpretations of statistics – even though the reported \textit{numeric} summaries are correct. We will also explain why it would be correct to instead have reported that\vspace{-0.470cm}
\blockquote{\normalsize “After HDPS adjustment for confounding, a 61\% hazard elevation remained; however, under the same model, every hypothesis from no elevation up to a 160\% hazard increase had $p$ > 0.05; Thus, while quite imprecise, these results are consistent with previous observations of a positive association between serotonergic antidepressant prescriptions and subsequent ASD. Because the association may be partially or wholly due to uncontrolled biases, further evidence will be needed for evaluating what, if any, proportion of it can be attributed to causal effects of prenatal serotonergic antidepressant use on ASD incidence."}\vspace{-0.40cm} 
We believe this type of language is careful and nuanced, and that such cautious attention to detail is essential for accurate scientific reporting. For simplicity and consistency with common practice we have used the 0.05 cutoff in the description, but recognize that researchers may be better served by choosing their descriptive approach as well as decision cutoffs based on background literature and error costs, rather than using traditional conventions \cite{Lakens2018-lg}. 
\vspace{-0.20cm}
\section{Making Sense of Tests}  
\vspace{-.750cm}
\noindent\textcolor{white}{\rule[-0.5ex]{\linewidth}{0.05pt}}
\vspace{-1.10cm}
\subsection{The \textit{P}-value as a \\compatibility measure}

The infamous \textit{observed} \textit{P}-value \textit{p} (originally called the observed or attained “level of significance” or “value of \textit{P}” \cite{fisherStatisticalMethodsResearch1925,pearsonCriterionThatGiven1900,stiglerAttemptsReviveBinomial1986}) is a measure of compatibility between the observed data and a targeted test hypothesis \textbf{H}, given a set of background assumptions (the background model) which are used along with the hypothesis to compute the \textit{P}-value from the data. By far the most common example of a test hypothesis \textbf{H} is a traditional null hypothesis, such as “there is no association” or (more ambitiously) “there is no treatment effect.” 

In some books this null hypothesis is the only test hypothesis ever mentioned. Nonetheless, the test hypothesis \textbf{H} could just as well be “the treatment doubles the risk” or “the treatment halves the risk” or any other hypothesis of practical interest \cite{greenlandStatisticalTestsValues2016}; we will argue such alternatives to the null \textit{should} also be tested whenever the traditional null hypothesis is tested. Our discussion will also apply when \textbf{H} concerns multiple parameters and thus the test involves multiple degrees of freedom, for example a general test of linearity of trend (dose-response) when a treatment has 5 levels (which has 3 degrees of freedom). 

With this general background about the test hypothesis, the other key ingredient in traditional statistical testing is a test statistic, such as a z-score or $\chi^{2}$, which measures the discrepancy between the observed data and what would have been expected under the test hypothesis, given the background assumptions. We can now define an observed \textit{P}-value \textit{p} as the probability of the test statistic being \textit{at least} as extreme as observed \textit{if} the hypothesis \textbf{H} targeted for testing \textit{and} every assumption used to compute the \textit{P}-value (the test hypothesis \textbf{H} \textit{and} the background statistical model) were correct \cite{greenlandStatisticalTestsValues2016}. Those background assumptions typically include a host of conditions such as linearity of responses and additivity of effects on a given scale; appropriateness of included variables (e.g., no intermediates for the effect under study); unimportance of omitted variables (e.g., all important confounding is controlled), random errors in a given family, no selection bias, and full accounting for measurement error and model selection.

This accurate and technical description does not accord well with human psychology, however: It is often said by Bayesians that researchers want a probability for the targeted test hypothesis (posterior probability of \textbf{H}), not a probability of observations. This imperative is indicated by the many “intuitive” – and incorrect – verbal definitions and descriptions of the \textit{P}-value that amount to calling it the probability of the test hypothesis, which is quite misleading \cite{greenlandStatisticalTestsValues2016}. Such errors are often called \textit{inversion fallacies} because they invert the role of the observations and the hypothesis in defining the \textit{P}-value (which is a probability for the observed test statistic, not the test hypothesis).

A standard frequentist criterion for judging whether a \textit{P}-value is valid for statistical testing is that all possible values for it from zero to one are equally likely (uniform in probability) if the test hypothesis and background assumptions are correct. We discuss this criterion in more detail in the \href{https://arxiv.org/abs/2008.12991}{Supplement} \cite{greenlandTechnicalIssuesInterpretation2020}. With this validity criterion met, we can also correctly describe the \textit{P}-value without explicit reference to repeated sampling, as the \textit{percentile} or proportion at which the observed test statistic falls in the distribution for the test statistic under the test hypothesis and the background assumptions \cite{perezgonzalezPvaluesPercentilesCommentary2015,vosFrequentistInferenceRepeated2019}. The purpose of this description is to connect the \textit{P}-value to a familiar concept, the percentile at which someone’s score fell on a standard test (e.g., a college or graduate admissions examination), as opposed to the remote abstraction of infinitely repeated sampling.
\vspace{-0.35cm}

\subsection{The \textit{S}-value}
{\fontsize{8.25}{13.25}\selectfont{
Even when \textit{P}-values are correctly defined and valid, their scaling can be deceptive due to their compression into the interval from 0 to 1, with vastly different meanings for absolute differences in \textit{P}-values near 1 and the same differences for \textit{P}-values near 0  \cite{greenlandValidPvaluesBehave2019}, as we will describe below. One way to reduce test misinterpretations and provide more intuitive numerical results is to translate the \textit{P}-values into probabilities of outcomes in familiar games of chance.  \\ \indent
Consider a game in which one coin will be tossed and we will bet on tails. Before playing however we want evidence that the tossing is acceptable for our bet, by which we mean not biased toward heads, because such loading would make our losing more probable than not. To check acceptability, suppose we first do \textit{s} independent test tosses and they \textit{all} come up heads. If the tossing is acceptable, the chance of this happening is at most ${\nicefrac{1}{2}}^s$, the chance of all heads in \textit{s} unbiased (fair) tosses. The smaller this chance, the less we would trust that the game is acceptable. In fact we could take s as measuring our evidence against acceptability: If we only did one toss and it came up heads (\textit{s} = 1) that would be unsurprising if the tossing were unbiased for then it would have chance ${\nicefrac{1}{2}}$, and so would provide barely any evidence against acceptability. But if we did 10 tosses and all came up heads (\textit{s} = 10) that would be surprising if the tossing were unbiased, for the chance of that is then ${\nicefrac{1}{2}}^{10}$ $\approx$ 0.001, and so would provide considerably more evidence against acceptability. \\ \indent
With this setting in mind, we can now gauge the evidence supplied by a \textit{P}-value \textit{p} by seeing what number \textit{s} of heads in a row would come closest to \textit{p}, which we can find by solving the equation \textit{p} = $\nicefrac{1}{2}^{s}$ for \textit{s}.  The solution is the negative base-2 logarithm of the \textit{P}-value, \textit{s} = $\log_{2}(\nicefrac{1}{p})$ = $−\log_{2}(p)$, known as the binary Shannon information, surprisal, logworth, or \textit{S}-value from the test \cite{Shannon1948-uq,greenlandValidPvaluesBehave2019,goodSurpriseIndexMultivariate1956, goodCorrectionsSurpriseIndex1957}. The \textit{S}-value is designed to reduce incorrect probabilistic interpretations of statistics by providing a nonprobability measure of information supplied by the test statistic against the test hypothesis \textbf{H} \cite{greenlandValidPvaluesBehave2019}. \\ \indent
The \textit{S}-value provides an absolute scale on which to view the information provided by a valid \textit{P}-value, as measured by calibrating the observed \textit{p} against a physical mechanism that produces data with known probabilities. A single coin toss produces a binary outcome which can be coded as 1 = heads, 0 = tails, and thus requires only two symbols or states to record or store; hence the information in a single toss is called \textit{bit}, short for binary digit, or a \textit{shannon}.  The information describing a sequence of \textit{s} tosses requires \textit{s} bits to record or store; thus, extending this measurement to a hypothesis \textbf{H} with \textit{P}-value \textit{p}, we say the test supplied \textit{s} = $−\log_{2}(p)$ bits of information against \textbf{H}. \\ \indent
We emphasize that, without further restrictions, our calibration of the \textit{P}-value against coin-tossing is only measuring information against the test hypothesis, not in support of it. This limitation is for the purely logical reason that there is no way to distinguish among the infinitude of background assumptions that lead to a test with the same or larger \textit{P}-value and hence the same or smaller \textit{S}-value. There is no way the data can \textit{support} a test hypothesis except relative to a fixed set of background assumptions. Rather than taking the background assumptions for granted, we prefer instead to adopt a refutational view, which emphasizes that any claim of support will be undermined by assumption uncertainty, and is thus best avoided. This caution applies regardless of the test statistic used, whether \textit{P}-value, \textit{S}-value, Bayes factor, or posterior probability. \\ \indent
As with the \textit{P}-value, the \textit{S}-value refers only to a particular test with particular background assumptions. A different test based on different background assumptions will usually produce a different \textit{P}-value and thus a different \textit{S}-value; thus it would be a mistake to simply call the \textit{S}-value “the information against the hypothesis supplied \textbf{by the data}”, for it is always a test of the hypothesis conjoined with (or conditioned on) the assumptions. As a basic example, we may contrast the \textit{P}-value for the strict null hypothesis (of no effect on any experimental unit) comparing two experimental groups using a $t$-test (which, along with randomization, assumes normally distributed responses under the null hypothesis), to the \textit{P}-value from a permutation test (which assumes only randomization). \\ \indent
Finally, as explained in the \href{https://arxiv.org/abs/2008.12991}{Supplement}, the \textit{S}-value can also be expressed using other logarithmic units such as natural (base-e) logs, $−\ln(p)$, which is mathematically more convenient but not as easy to represent physically.}
\vspace{-0.70cm}
}

\begin{figure*}[t]
\begin{center}
\fbox{\includegraphics[width=27.5pc, height=21pc]{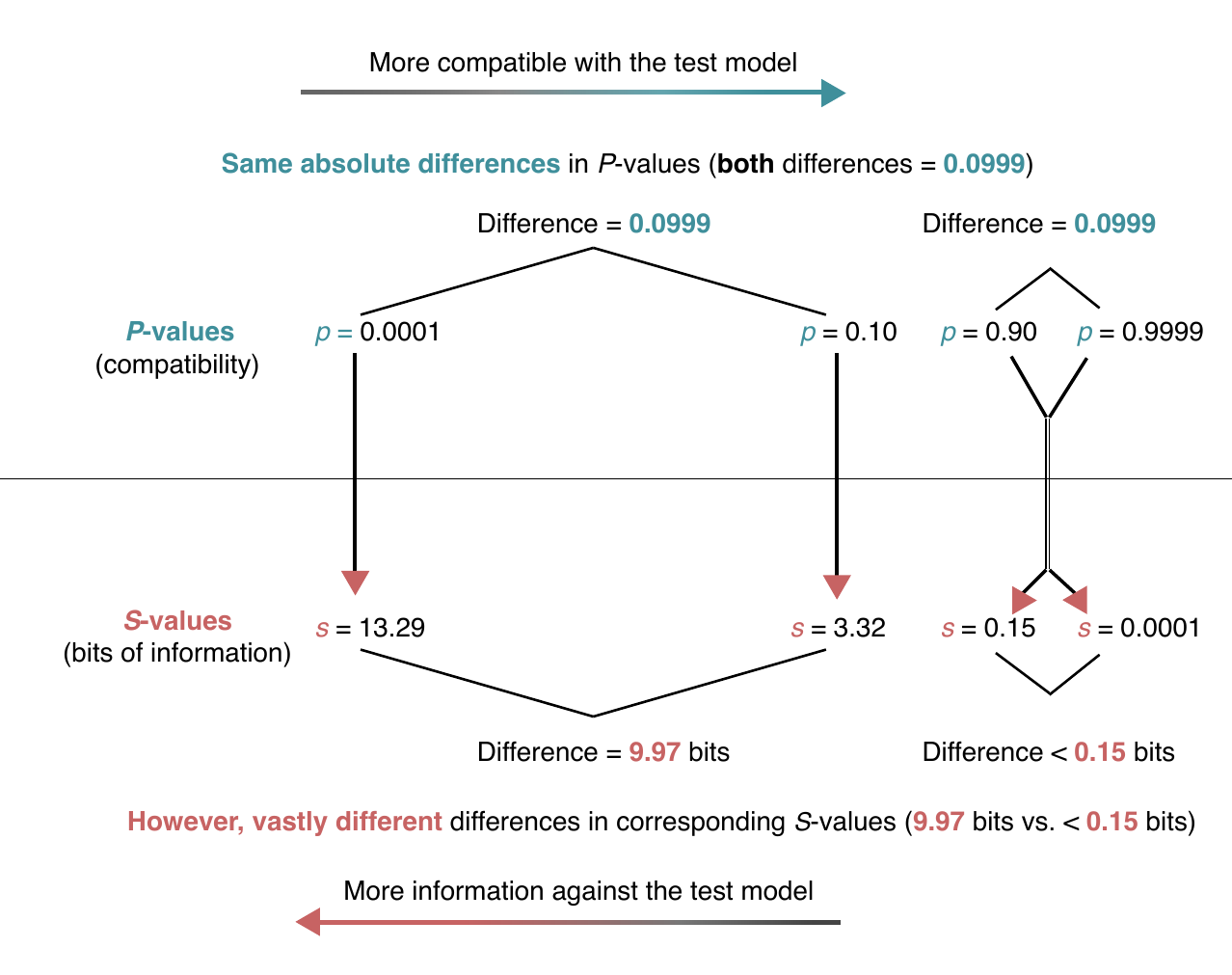}}
\captionsetup{width=.75\linewidth}
\caption{\small\textbf{Comparison of \textit{P}-value and \textit{S}-value scales.} \textbf{Top labels:} Data compatibility with test model as measured by \textit{P}-values. \textbf{Bottom labels:} Information against test model as measured by the corresponding \textit{S}-values.}
\label{Fig1}
\vspace{-0.25cm}
\begin{adjustwidth*}{50pt}{50pt}
\noindent\textcolor{gray!50}{\rule[-0.5ex]{\linewidth}{0.05pt}}
\end{adjustwidth*}
\vspace{-1cm}
\end{center}
\end{figure*}
\vspace{-0.25cm}

\subsection[Evaluating \textit{P}-values and fixed-cutoff tests with \textit{S}-values]{\texorpdfstring{Evaluating \textit{P}-values and \\ fixed-cutoff tests with \\ \textit{S}-values}{Evaluating \textit{P}-values and fixed-cutoff tests with \textit{S}-values}}

{\fontsize{8.25}{20}\selectfont{
With the \textit{S}-value in hand, a cognitive difficulty of the \textit{P}-value scale for evidence can be seen by first noting that the difference in the evidence provided by \textit{P}-values of 0.9999 and 0.90 is trivial: Both represent almost no information against the test hypothesis, in that the corresponding \textit{S}-values are $−\log_{2}(0.9999)$ =  0.00014 bits and $−\log_{2}(0.90)$ = 0.15 bits. Both are far less than 1 bit of information against the hypothesis – they are just a fraction of a coin toss different. In contrast, the information against the test hypothesis in \textit{P}-values of 0.10 and 0.0001 is profoundly different, in that the corresponding \textit{S}-values are $−\log_{2}(0.10)$ = 3.32 and $−\log_{2}(0.0001)$ = 13.3 bits; thus \textit{p} = 0.001 provides 10 bits more information against the test hypothesis than does \textit{p} = 0.10, corresponding to the information provided by 10 additional heads in a row. The contrast is illustrated in \textbf{Figure \ref{Fig1}}, along with other examples of the scaling difference between \textit{P} and \textit{S} values.  \\ \indent
As an example of this perspective on reported results, from the point and interval estimate from the HDPS analysis reported by Brown et al. \cite{brownAssociationSerotonergicAntidepressant2017}, we calculated that the \textit{P}-value for the “null” test hypothesis \textbf{H} that the hazard ratio is 1 (no association) is 0.0505. Using the \textit{S}-value to measure the information supplied by the HDPS analysis against this hypothesis, we get \textit{s} = $−\log_{2}(0.0505)$ = 4.31 bits; this is hardly more than 4 coin tosses worth of information against no association. For comparison, when setting the test hypothesis \textbf{H} to be that the hazard ratio is 2 (doubling of the hazard among the treated), we calculated a \textit{P}-value of about 0.373. \\ \indent
The information supplied by the HDPS analysis against this test hypothesis is then measured by the \textit{S}-value as \textit{s} = $−\log_{2}(0.373)$ = 1.42 bits, hardly more than a coin-toss worth of information against doubling of the hazard among the treated. In these terms, then, the HDPS results supply roughly 3 bits more information against no association than against doubling of the hazard, so that doubling (HR = 2) is more compatible with the analysis results than is no association (HR = 1).\\ \indent
\textit{S}-values help clarify objections to comparing \textit{P}-values to sharp dichotomies. Consider that a \textit{P}-value of 0.06 yields about 4 bits of information against the test hypothesis \textbf{H}, while a \textit{P}-value of 0.03 yields about 5 bits of information against \textbf{H}. Thus, \textit{p} = 0.06 is about as surprising as getting all heads on four fair coin tosses while \textit{p} = 0.03 is one toss (one bit) more surprising. Even if one is committed to making a decision based on a sharp cutoff, \textit{S}-values illustrate what range around that cutoff corresponds to a trivial information difference (e.g., any \textit{P}-value between 0.025 and 0.10 is less than a coin-toss difference in evidence from \textit{p} = 0.05).}} 
\end{multicols}
 
\newpage
\vspace{-0.25cm}
\begin{adjustwidth*}{55pt}{55pt}
\indent

{\fontsize{8.5}{20}\selectfont{
\textit{S}-values also help researchers understand more subtle problems with traditional testing. Consider for example the import of the magical 0.05 threshold ($\alpha$-level) that is used to declare associations present or absent. It has often been claimed that this threshold is too high to be regarded as representing much evidence against \textbf{H} \cite{Benjamin2017-fz,sellkeCalibrationValuesTesting2001}, but the arguments for that are usually couched in Bayesian terms of which many remain skeptical. We can however see those objections to 0.05 straightforwardly by noting that the threshold translates into requiring an \textit{S}-value of only $−\log_{2}(0.05)$ = 4.32 bits of information against the null; that means \textit{p} = 0.05 is barely more surprising than getting all heads on 4 fair coin tosses. \\ \indent
While 4 heads in a row may seem surprising to some intuitions, it does in fact correspond to doing only 4 tosses to study the coin; a sample size of \textit{N} = 4 binary outcomes would rarely qualify as a basis for (say) recommending a new treatment even if all 4 patients recovered but the recovery rate without the treatment was known to be 50\%. Thus, like other proposals, the \textit{S}-value calls into question the traditional $\alpha$ = 0.05 standard, and may help users realize how little information is contained in most \textit{P}-values when compared to the thousands of bits of information in a typical cell-phone directory.}} \\
\vspace{.10cm}

\noindent\textcolor{gray!50}{\rule[-0.5ex]{\linewidth}{0.05pt}}
\end{adjustwidth*}

\newpage
\begin{adjustwidth*}{65pt}{65pt}
\noindent\textcolor{gray!50}{\rule[-0.5ex]{\linewidth}{0.05pt}}
\vspace{.40cm}

{\fontsize{8.5}{16}\selectfont{
Further crucial information will be given by \textit{P}-values and \textit{S}-values tabulated for several \textit{alternative} hypotheses, interval estimates over varying percentiles, and graphs of data and information summaries such as those illustrated below.

\section{Further advantages of \textit{S}-values}
\vspace{-.50cm}
\noindent\textcolor{gray!50}{\rule[-0.5ex]{\linewidth}{0.05pt}} \\ 
\vspace{.05cm}

Unlike probabilities, \textit{S}-values are unbounded above and can be added over independent information sources to create simple summary tests [55 p. 80; see also our \href{https://arxiv.org/abs/2008.12991}{Supplement} \cite{greenlandTechnicalIssuesInterpretation2020}]. They thus provide a scale for comparing and combining test results across studies that is aligned with information rather than probability measurement \cite{greenlandValidPvaluesBehave2019}. Another advantage of \textit{S}-values is that they help thwart inversion fallacies, in which a \textit{P}-value is misinterpreted as a probability of a hypothesis being correct (or equivalently, as the probability that a statement about the hypothesis is in error). Hypothesis probabilities computed using the data are called \textit{posterior probabilities} (because they come after the data). It is difficult to confuse an \textit{S}-value with a posterior probability because the \textit{S}-value is unbounded above, and in fact will be above 1 whenever the \textit{P}-value is below 0.50. \\

Probabilities of data summaries (test statistics) given hypotheses and probabilities of hypotheses given data are identically scaled, leading users to inevitably conflate \textit{P}-values with posterior probabilities. This confusion dominates observed misinterpretations \cite{greenlandStatisticalTestsValues2016} and is invited with open arms by “significance” and “confidence” terminology. Such mistakes could potentially be avoided by giving actual posterior probabilities along with \textit{P}-values. Bayesian methods provide such probabilities but require prior distributions as input; in turn, those priors require justifications based on often contentious background information. While the task of creating such distributions can be instructive, this extra input burden has greatly deterred adoption of Bayesian methods; in contrast, \textit{S}-values provide a direct quantification of information without this input. 

}}
\vspace{0.25cm}
\noindent\textcolor{gray!50}{\rule[-0.5ex]{\linewidth}{0.05pt}}

\end{adjustwidth*}

\newpage

\newgeometry{paperwidth=171mm,paperheight=200mm,
  left=24.3mm,right=24.3mm,textwidth=142.4mm,lines=47,
  headsep=\baselineskip,top=20mm,
  bottom=20mm}
  \fancyhfoffset[E,O]{0pt}
\afterpage{%

\noindent\textcolor{white}{\rule[-0.5ex]{\linewidth}{0.05pt}}
 \begingroup
 \def\arraystretch{1.1}%
 \small
 
 \vspace{-0.55cm}
 
 \captionof{table}{\textbf{\textit{P}-values and binary \textit{S}-values, with corresponding maximum-likelihood ratios (MLR) and deviance (likelihood-ratio) statistics for a simple test hypothesis \textbf{H} under background assumptions \textbf{A}.}}
 
 \vspace{-0.2cm}
\begin{center}
\setlength{\tabcolsep}{0mm}
\begin{tabular}[]{c<{\hspace{10mm}}c<{\hspace{-6mm}}c<{\hspace{6mm}}c<{\hspace{7mm}}c<{\hspace{7mm}}}\toprule\addlinespace[0.0em]
\headrow
  \thead{\small \textit{P}-value \textit{p} \\ {\scriptsize\hspace{1mm} (compatibility of H with data given A)}} & & \thead{\small \textit{S}-value \textit{s} = \textbf{-log$_{2}$(\textit{p})} \\ {\scriptsize (information against \textbf{H} given \textbf{A} in bits)}} & \thead{\small MLR \\ \scriptsize (against \textbf{H} given \textbf{A}}) & \thead{\small Deviance Statistic \\ {\scriptsize 2ln(MLR)}} \\ \midrule
   0.99                            & & 0.014          & 1.00                & 0.00016                \\
   0.90                             & & 0.15          & 1.01                & 0.016                  \\
    0.50                           &   & 1.00          & 1.26                & 0.45                   \\
    0.25                            &  & 2.00          & 1.94                & 1.32                   \\
    0.10                            &  & 3.32          & 3.87                & 2.71                   \\
    0.05                            &  & 4.32          & 6.83                & 3.84                   \\
    0.025                           &  & 5.32          & 12.3                & 5.02                   \\
    0.01                            &  & 6.64          & 27.6                & 6.63                   \\
    0.005                           &  & 7.64          & 51.4                & 7.88                   \\
  0.0001                           & & 13.3          & 1935                & 15.1                   \\
   5 sigma\protect\footnotemark ($\approx$ 2.9 in 10 million)            &  & 21.7          & $5.2\times10^{5}$   & 26.3                   \\
    1 in 100 million (GWAS)          &  & 26.6         & $1.4\times10^{7}$   & 32.8                  \\
   6 sigma\protect\footnotemark ($\approx$ 1 in a billion)         &  & 29.9          & $1.3\times10^{8}$   & 37.4                   \\ \bottomrule
 \end{tabular}
  \end{center}
 \label{Tab:tab1}
 \endgroup
  \vspace{-.2cm}
\footnotetext{\scriptsize 5 and 6 sigma cutoffs are the upper standard-normal tail probabilities at 5 and 6 standard\\ deviations above the mean \cite{cousinsJeffreysLindleyParadox2017}. Further details of the relations among these measures\\ are given in the \hyperref[sec:Appendix]{Appendix}.}

\begin{adjustwidth*}{25pt}{25pt} \indent

{\fontsize{8.25}{17}\selectfont{
\textbf{Table~\ref{Tab:tab1}} provides a translation of the \textit{P}-value to the binary \textit{S}-value $s$ = $−\log_{2}(p)$. It also gives the maximum-likelihood ratio (MLR), and the deviance-difference or likelihood-ratio statistic $2\ln(MLR)$, assuming that \textbf{H} is a simple hypothesis about one parameter (e.g., that a mean difference or regression coefficient is zero) and that the statistic has a 1 degree of freedom $\chi^{2}$ distribution; see the \hyperref[sec:Appendix]{Appendix} and \href{https://arxiv.org/abs/2008.12991}{Supplement} \cite{greenlandTechnicalIssuesInterpretation2020} for further details. The MLR and deviance statistic are themselves often treated as measures of information against \textbf{H} under the background assumptions (fortuitously, when rounding to the nearest integer, the binary \textit{S}-value and deviance statistic coincide in the often-contentious \textit{P}-value range of 0.005 to 0.10).}}\\ \indent
{\fontsize{8.25}{17}\selectfont{
The table also shows the different alpha levels used in various fields and the stark contrast in information associated with these cutoffs. The alpha levels used in particle physics and genome-wide association studies (GWAS) are extremely small because in those areas false positives are considered far more likely and costly than false negatives: Discovery declarations in particle physics require “5 sigmas”, nearly 22 bits of information against \textbf{H} (corresponding to all heads in 22 fair coin tosses), while GWAS requires nearly 27 bits; for discussions of these choices see \cite{cousinsJeffreysLindleyParadox2017, dudbridgeEstimationSignificanceThresholds2008}.}}\\ \indent
{\fontsize{8.25}{17}\selectfont{
Further details of the relations among these and other measures are given in the \href{https://arxiv.org/abs/2008.12991}{Supplement} \cite{greenlandTechnicalIssuesInterpretation2020}. \textbf{Table~\ref{Tab:tab2}} presents these measures as computed from the Brown et al. report  \cite{brownAssociationSerotonergicAntidepressant2017}; it can again be seen that by any measure there is more information against the null (equal hazards across treatment, \textit{S} = 4.31) than against doubling of the hazard (HR = 2, \textit{S} = 1.42), so the claim that these results demonstrate or support no association is simply wrong. }} 

\end{adjustwidth*}
 
\aftergroup\restoregeometry
}

\restoregeometry

\newpage
\vspace{.5cm}

\begin{center}
\begingroup
\def\arraystretch{1.75}%
\small
\vspace{.75cm}
\captionof{table}{\textbf{\textit{P}-values, \textit{S}-values, Maximum-Likelihood Ratios, and Likelihood-Ratio Statistics For Various Test Hypotheses About the Hazard Ratio (HR) Computed from Brown et al. \cite{brownAssociationSerotonergicAntidepressant2017} HDPS results.\protect\footnotemark}}

\vspace{.35cm}

\setlength{\tabcolsep}{0mm}
\begin{tabular}[]{l<{\hspace{10mm}}c<{\hspace{-10mm}}c<{\hspace{10mm}}c<{\hspace{5mm}}c<{\hspace{5mm}}c<{\hspace{5mm}}}\toprule\addlinespace[0.0em]
\headrow
\thead{Test Hypothesis (H)} & & \thead{\textit{P}-value \\ (compatibility)}  & \thead{\textit{S}-value \\ (bits of information)} & \thead{Maximum-\\Likelihood Ratio} & \thead{Likelihood-\\Ratio Statistic} \\ \midrule
Halving of hazard (HR = 0.5) &       & $1.6\times10^{-6}$      & 19.3      & $1.0\times10^{5}$     & 23.1     \\
No association (null) (HR = 1) &     & 0.05                    & 4.31      & 6.77                  & 3.82     \\
Point estimate (HR = 1.61) &         & 1                       & 0.00      & 1.00                  & 0.00     \\
Doubling of hazard (HR = 2) &        & 0.37                    & 1.42      & 1.49                  & 0.79     \\
Tripling of hazard (HR = 3) &        & 0.01                    & 6.56      & 26.2                  & 6.53     \\
Quintupling of hazard (HR = 5) &     & $3.3\times10^{-6}$      & 18.2      & $5.0\times10^{4}$     & 21.7     \\ \bottomrule
\end{tabular}
\label{Tab:tab2}
\endgroup
\end{center}
{\color{black!100}\footnotetext{Computed from the normal approximations given in the \hyperref[sec:Appendix]{Appendix}.}}

\normalsize 

\begin{adjustwidth*}{60pt}{60pt}
\indent
{\fontsize{8.75}{16.5}\selectfont{
\vspace{.10cm}

In summary, the \textit{S}-value provides a gauge of the information supplied by a statistical test in familiar terms of coin tosses. It thus complements the probability interpretation of a \textit{P}-value by supplying a mechanism that can be visualized with simple physical experiments. Given amply documented human tendencies to underestimate the frequency of seemingly “unusual” events \cite{handImprobabilityPrincipleWhy2014}, these experiments can guide intuitions about what evidence strength a given \textit{P}-value actually represents. }}

\noindent\textcolor{gray!50}{\rule[-0.5ex]{\linewidth}{0.05pt}}
\vspace{-.850cm}
\section{Replace Unrealistic Confidence Claims with Compatibility Measures}
{\fontsize{8.75}{15.5}\selectfont{
Confidence intervals (commonly abbreviated as CI) have been widely promoted as a solution to the problems of statistical misinterpretation \cite{cummingUnderstandingNewStatistics2012,rothmanShowConfidence1978}. While we support their presentation, such intervals have difficulties of their own.  The major problem with “confidence” is that it encourages the common confusion of the CI percentile level (typically 95\%) with the probability that the true value of the parameter is in the interval (mistaking the CI for a Bayesian posterior interval) \cite{greenlandStatisticalTestsValues2016},  as in statements such as “we are 95\% \textit{confident} that the true value is within the interval.” \\ \indent
The fact that “confidence” refers to the procedure behavior, \textit{not} the reported interval, seems to be lost on most researchers. Remarking on this subtlety, when Jerzy Neyman discussed his confidence concept in 1934 at a meeting of the Royal Statistical Society, Arthur Bowley replied, “I am not at all sure that the 'confidence' is not a confidence trick.” \cite{bowleyDiscussionDrNeyman1934}.}} \\
\vspace{.60cm}
\end{adjustwidth*}

\newpage
\begin{adjustwidth*}{50pt}{50pt}

\noindent\textcolor{gray!50}{\rule[-0.5ex]{\linewidth}{0.05pt}}

\vspace{.60cm}

{\fontsize{8.75}{20}\selectfont{
And indeed, forty years later, Cox and Hinkley warned, “interval estimates cannot be taken as probability statements about parameters, and foremost is the interpretation ‘such and such parameter values are consistent with the data.’ ” \cite{coxChapterIntervalEstimation1974}. Unfortunately, the word “consistency” is used for several other concepts in statistics, while in logic it refers to an absolute condition (of noncontradiction); thus, its use in place of “confidence” would risk further confusion. \\ \indent
 To address the problems above, we exploit the fact that a 95\% CI summarizes the results of varying the test hypothesis \textbf{H} over a range of parameter values, displaying all values for which  \textit{p} > 0.05 \cite{coxPrinciplesStatisticalInference2006} and hence \textit{s} < 4.32 bits \cite{amrheinInferentialStatisticsDescriptive2019,greenlandValidPvaluesBehave2019}. Thus the CI contains a range of parameter values that are more compatible with the data than are values outside the interval, under the background assumptions \cite{greenlandValidPvaluesBehave2019,greenlandStatisticalTestsValues2016}. Unconditionally (and thus even if the background assumptions are uncertain), the interval shows the values of the parameter which, when combined with the background assumptions, produce a test model that is “highly compatible” with the data in the sense of having less than 4.32 bits of information against it. We thus refer to CI as \textit{compatibility} intervals rather than \textit{confidence} intervals  \cite{amrheinInferentialStatisticsDescriptive2019,greenlandAreConfidenceIntervals2019,greenlandValidPvaluesBehave2019}; their abbreviation remains “CI.” We reject calling these intervals “uncertainty intervals,” because they do \textbf{not} capture uncertainty about background assumptions \cite{greenlandAreConfidenceIntervals2019}. \\ \indent
Another problem is that a frequentist CI is often used as nothing more than a null-hypothesis significance test (NHST), by declaring that the null parameter value (e.g., HR = 1) is supported if it is inside the interval, or refuted if it is outside the interval. These declarations defeat the use of interval estimates to summarize information about the parameter, and perpetuate the fallacy that information changes abruptly across decision boundaries \cite{amrheinInferentialStatisticsDescriptive2019,greenlandStatisticalTestsValues2016,pooleConfidenceIntervalsExclude1987,Poole1987-nb}. In particular, the usual 95\% default forces the user’s focus onto parameter values that yield \textit{p} > 0.05, without regard to the trivial difference between (say) \textit{p} = 0.06 and \textit{p} = 0.04 (an information difference far smaller than a coin toss). Even differences conventionally seen as “large” are often minor in information terms, e.g., \textit{p} = 0.02 and \textit{p} = 0.16 represent a difference of only $\log_{2}(\frac{0.16}{0.02})$ = 3 coin tosses, underscoring the caution that the difference between “significance” and “nonsignificance” is not significant\cite{gelmanDifferenceSignificantNot2006}. \\ \indent
To address this problem, we first note that a 95\% interval estimate is only one of a number of arbitrary dichotomization of possibilities of parameter values (into either inside or outside of an interval). A more accurate picture of information is then obtained by examining intervals using other percentiles, e.g., proportionally-spaced compatibility levels such as \textit{p} > 0.25, 0.05, 0.01, which correspond to 75\%, 95\%, 99\% CIs and equally-spaced \textit{S}-values of \textit{s} < 2, 4.32, 6.64 bits. When a detailed picture is desired, a table or graph of \textit{P}-values and \textit{S}-values across a broad range of parameter values seems the clearest way to see how compatibility varies smoothly across the values.}} \\ 
\vspace{.20cm}
\noindent\textcolor{gray!50}{\rule[-0.5ex]{\linewidth}{0.05pt}}
\end{adjustwidth*}

\newpage

\begin{adjustwidth*}{60pt}{60pt}

\begin{figure*}
\begin{center}
\fbox{\includegraphics[width=30pc, height=18pc]{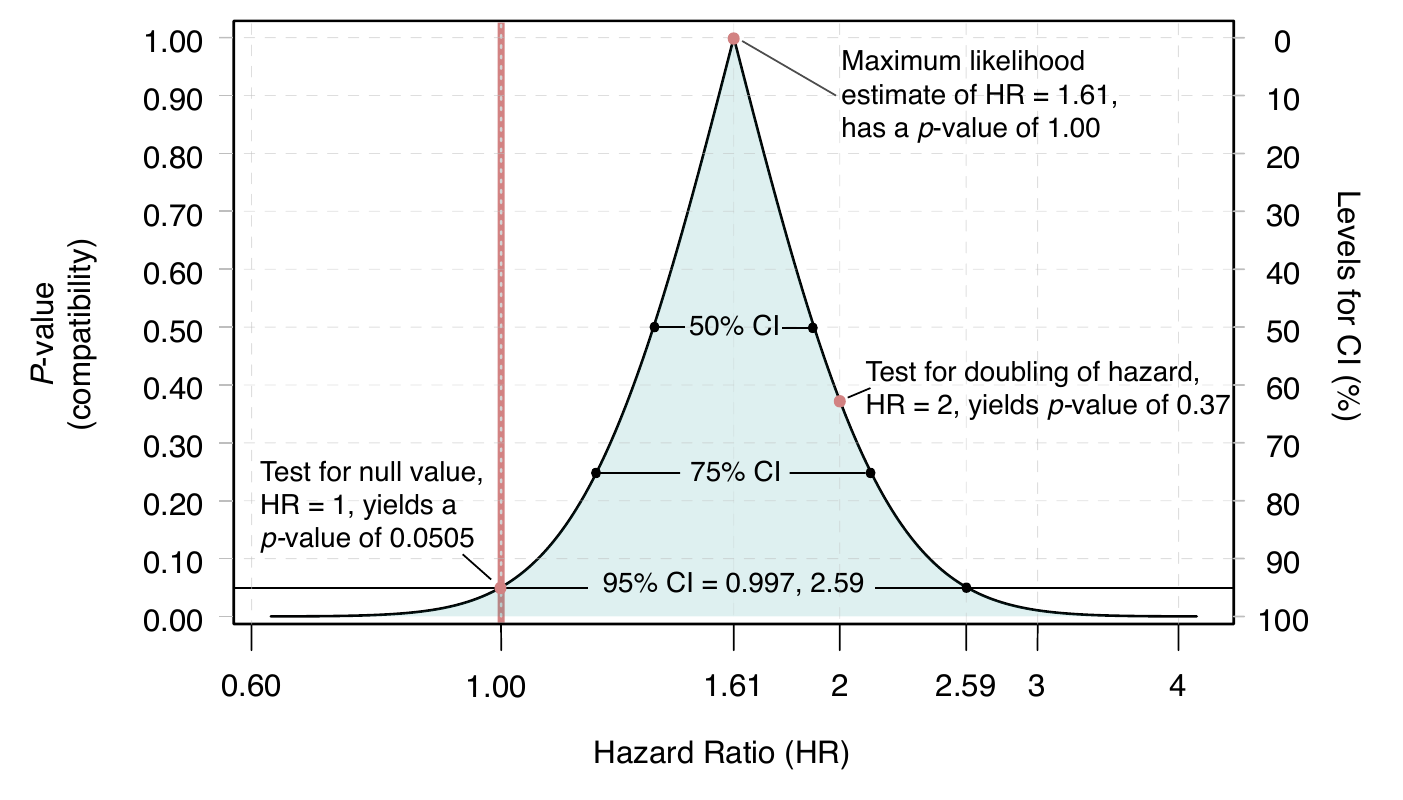}}
\captionsetup{width=.90\linewidth}
\caption{\small\textbf{\textit{P}-values for a range of hazard ratios (HR).} A compatibility graph in which \textit{P}-values are plotted across alternative hazard ratios. Computed from results in Brown et al. \cite{brownAssociationSerotonergicAntidepressant2017}. Compatibility intervals (CI) in percents can be read moving from the right-hand axis to the bottom (HR) axis. HR = 1 represents no association.}
\label{Fig2}
\begin{adjustwidth*}{60pt}{60pt}
\noindent\textcolor{white}{\rule[-0.5ex]{\linewidth}{0.05pt}}
\end{adjustwidth*}
\vspace{-.850cm}
\end{center}
\end{figure*}
\end{adjustwidth*}

\begin{adjustwidth*}{55pt}{55pt}
\section{Gradations, Not Dichotomies}
\vspace{-.150cm}

{\fontsize{8}{18}\selectfont{
Graphs of \textit{P}-values or their equivalent have been promoted for decades \cite{birnbaumUnifiedTheoryEstimation1961,Poole1987-nb,rothmanPrecisionStatisticsEpidemiologic2008,Sullivan1990-ha}, yet their adoption has been slight. Nonetheless, \textit{P}-value and \textit{S}-value graphing software is now available freely through several statistical packages \cite{rafiConcurveComputesPlots2020,ruckerForestPlotDrapery2020}. A graph of the \textit{P}-values \textit{p} against possible parameter values allows one to see at a glance which parameter values are most compatible with the data under the background assumptions. This graph is known as the \textit{P}-value function, or compatibility, consonance, or confidence curve \cite{birnbaumUnifiedTheoryEstimation1961,Poole1987-nb,rothmanPrecisionStatisticsEpidemiologic2008,Sullivan1990-ha,fraserPvalueFunctionStatistical2019,Singh2007-zr,xieConfidenceDistributionFrequentist2013,whiteheadCaseFrequentismClinical1993,schwederConfidenceLikelihoodProbability2016}; the “severity curve” (\cite{mayoStatisticalInferenceSevere2018}, Fig. 5.5) is a special case (see our \href{https://arxiv.org/abs/2008.12991}{Supplement} \cite{greenlandTechnicalIssuesInterpretation2020}). Transforming the corresponding \textit{P}-values in the graph to \textit{S}-values produces an \textit{S}-value (surprisal) function. \\ \indent 
Most studies not only examine but also present results for multiple associations and models, and examining or presenting graphs for each of the results may be impractical. Nonetheless, as in the Brown et al. example, there is often a “final” analysis or set of results that is used to generate the highlighted conclusions of the study. We strongly advise inspecting graphs for those analyses before writing conclusions, and presenting the graphs in the paper or at least in a supplementary file. As mentioned above, it is quite easy to now construct these curves using various statistical packages \cite{rafiConcurveComputesPlots2020,ruckerForestPlotDrapery2020}. }}

{\fontsize{8}{18}\selectfont{
\textbf{Figure \ref{Fig2}} and \textbf{Figure \ref{Fig3}} give the \textit{P}-value and \textit{S}-value graphs produced from the Brown et al. \cite{brownAssociationSerotonergicAntidepressant2017} data, displaying an estimated hazard ratio of 1.61 and 95\% limits of 0.997, 2.59 (see \hyperref[sec:Appendix]{Appendix} for computational details). Following the common (and important) warning that \textit{P}-values are not hypothesis probabilities, we caution that the \textit{P}-value graph is not a probability distribution: It shows compatibility of parameter values with the data, rather than plausibility or probability of those values given the data. This is not a subtle difference: compatibility is a much weaker condition than plausibility. Consider for example that complete fabrication of the data is always an explanation \textit{compatible} with the data (and indeed has happened in some influential medical studies \cite{rubensteinNewLowDrug2009}), but in studies with many participants and authors involved in all aspects of data collection it becomes so implausible as to not even merit mention. We emphasize then that all the \textit{P}-value ever addresses in a direct logical sense is compatibility; for hypothesis probabilities one must turn to Bayesian methods \cite{greenlandValidPvaluesBehave2019}.}}

\begin{figure*}
\begin{center}
\fbox{\includegraphics[width=28pc, height=18pc]{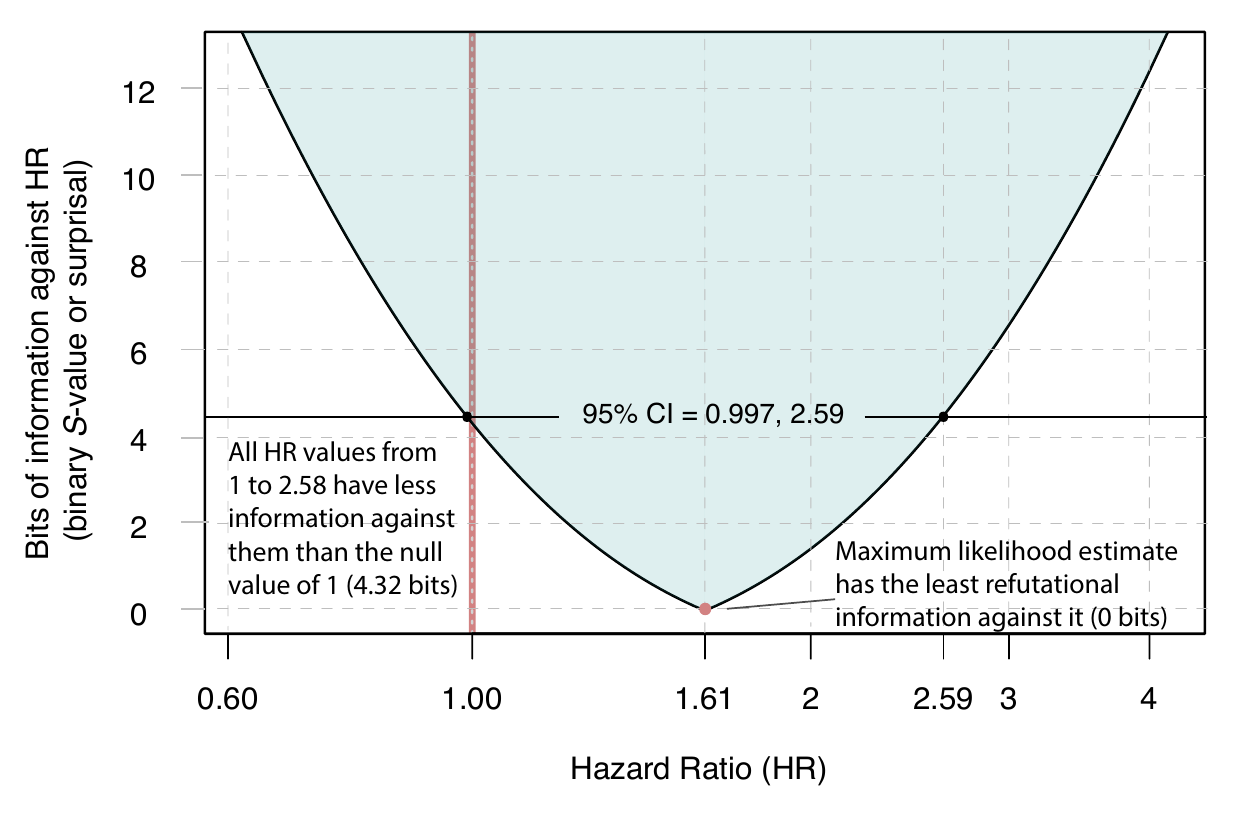}}
\captionsetup{width=.80\linewidth}
\caption{\small\textbf{\textit{S}-values (surprisals) for a range of hazard ratios.} An information graph in which \textit{S}-values are plotted across alternative hazard ratios. Computed from results in Brown et al. \cite{brownAssociationSerotonergicAntidepressant2017}. HR = 1 represents no association.}
\label{Fig3}
\begin{adjustwidth*}{60pt}{60pt}
\noindent\textcolor{gray!50}{\rule[-0.5ex]{\linewidth}{0.05pt}}
\end{adjustwidth*}
\vspace{-1.45cm}
\end{center}
\end{figure*}

{\fontsize{8}{18}\selectfont{
The \textit{P}-value graph rises past HR = 1 (no association, a parameter value which we have only plotted for demonstration purposes) until it peaks at the point estimate of 1.61, which coincides with the smallest \textit{S}-value. The graphs show how rapidly the \textit{P}-values fall and the \textit{S}-values rise as we move away from the point estimate. CIs at the 75, 95, and 99 percent levels can be read off the graph as the range between the parameter values where the graph is above the \textit{P} = 0.25, 0.05, and 0.01 levels. Both \textbf{Figure \ref{Fig2}} and \textbf{Figure \ref{Fig3}} illustrate how misleading it is to frame discussion in terms of whether \textit{P} is above or below 0.05, or whether the null value is included in the 95\% CI: Every hazard ratio from 1 to 2.58 is more compatible with the Brown et al. \cite{brownAssociationSerotonergicAntidepressant2017} data according to the HDPS analysis, and has less information against it than does the null value of 1. Thus, the graphs illustrate how the Brown et al. analysis provides absolutely no basis for claiming the study found “no association.” Instead, their analysis exhibits an association similar to that seen in earlier studies and should have been reported as such, even though it leaves open the question of what \textit{caused} the association (e.g., a drug effect, a bias, a positive random error, or some combination) and whether a clinically important effect is present.}} 

\end{adjustwidth*}

\begin{multicols}{2} 
\section{Discussion}
We now discuss several basic issues in the use of the methods we have described. The \href{https://arxiv.org/abs/2008.12991}{Supplement} \cite{greenlandTechnicalIssuesInterpretation2020} discusses several more technical topics mentioned earlier and below: Different units for the \textit{S}-value besides base-2 logs (bits); the importance of uniformity (validity) of the \textit{P}-value for interpretation of the \textit{S}-value; and the relation of the \textit{S}-value to other measures of statistical information about a test hypothesis or model.
\vspace{-.750cm}
\noindent\textcolor{white}{\rule[-0.5ex]{\linewidth}{0.05pt}}
\vspace{-.50cm}
\subsection{Moving Forward}

Most efforts to reform statistical reporting have promoted interval estimates \cite{cummingUnderstandingNewStatistics2012,rothmanShowConfidence1978} or Bayesian methods \cite{sellkeCalibrationValuesTesting2001} over \textit{P}-values. There is nonetheless scant empirical evidence that these or any proposals (including ours) have improved or will improve reporting without accompanying editorial and reviewer efforts to enforce proper interpretations. Instead, the above example and many others \cite{greenlandValidPvaluesBehave2019,greenlandSeriousMisinterpretationConsistent2017,schmidtMistakenInferenceCaused2014} illustrate how, without proper editorial monitoring, interval estimates are often of no help and can even be harmful when journals require dichotomous interpretation of results, for example as does \textit{JAMA} \cite{bauchnerReportingInterpretationRandomized2019}. \\ \indent
Cognitive psychology and its offshoot of behavioral economics (the “heuristics and biases” literature) have been studying misperceptions of probability for at least a half-century (e.g., see the anthologies of \cite{gilovichHeuristicsBiasesPsychology2002,tverskyJudgmentUncertaintyHeuristics1974}), with increasing attention to the harms of null-hypothesis significance testing (e.g.,  \cite{gigerenzerMindlessStatistics2004,gigerenzerSurrogateScienceIdol2015}). Informal classroom observations on the devices we discuss have been encouraging (both our own and those reported to us anecdotally by colleagues), leading to the present exposition.  \\ \indent
We would thus encourage formal experiments to study cognitive devices like those we discuss. To justify such effort, the devices must be well-grounded in statistical theory (as reviewed in the prequel to this article \cite{greenlandValidPvaluesBehave2019}), and should be clearly operationalized, as the current article attempts to do. These preliminaries are especially important because prevailing practice is cemented into nearly a century of routine teaching and journal demands; thus, any comparison today will be dramatically confounded by tradition and exposure. Addressing this imbalance will require detailed instruction in the graphical information perspective, as illustrated here.

\vspace{-.950cm}
\noindent\textcolor{white}{\rule[-0.5ex]{\linewidth}{0.05pt}}
\vspace{-.850cm}
\subsection{Tests of model fit}
\vspace{-.50cm}
For simplicity we have focused on tests of specific hypotheses given a set of assumptions (the background model). The \textit{S}-value can also be used to measure information against a data model, as supplied by the \textit{P}-value from general tests of fit of a model to the data (such as the Pearson \cite{pearsonCriterionThatGiven1900} chi-squared test of fit). In those tests, all deviations of the data from the model predictions contribute to lack of fit and are cumulated as evidence against the model. In yet another unfortunate misnaming, these tests have come to be called “goodness of fit” tests, when in fact the test statistics are measuring misfit (in Pearson’s case, squared distances between the predictions and observations). The \textit{P}-value accounts for the residual degrees of freedom for the misfit, but as discussed before, is scaled in a nonintuitive way: It shrinks to zero as misfit increases, even when misfit can increase indefinitely. The \textit{S}-value restores the proper relation to the fit as seen in the original test statistic, where the cumulative information against the model growing larger without bound as misfit increases without bound.
\vspace{-2cm} 
\noindent\textcolor{white}{\rule[-0.5ex]{\linewidth}{0.05pt}}
\vspace{.50cm} 
\subsection{Connections to Bayesian and information statistics} 
\vspace{-1.0cm}
\noindent\textcolor{white}{\rule[-0.5ex]{\linewidth}{0.05pt}}\vspace{.25cm} \\
Our development has been based on conventional frequentist statistics, which focus on probabilities of various statistical observations (data features). There are several connections of \textit{P}-values and compatibility intervals to Bayesian statistics, which are expressed in terms of hypothesis probabilities; for a basic review see \cite{Greenland2013-oa}. These in turn lead to connections to \textit{S}-values. Consider for example a one-sided \textit{P}-value \textit{p} for a directional hypothesis; under certain assumptions \textit{p} is a lower bound on the posterior probability that the hypothesis is false, and the \textit{S}-value \textit{s} = $−\log_{2}(p)$ can be interpreted as the maximum surprisal in finding the hypothesis is false, given the data and assumptions. The \href{https://arxiv.org/abs/2008.12991}{Supplement} \cite{greenlandTechnicalIssuesInterpretation2020} describes a connection to Bayes factors, “safe testing”, and testing by betting scores. \\ \indent

\subsection{Some cautions}

Demands for more statistical evidence against test hypotheses increase the need for numerical accuracy, especially because traditional normal z-score (Wald) approximations (used by most software to derive \textit{P}-values and compatibility intervals under nonlinear models) deteriorate as the \textit{P}-value or $\alpha$-level becomes smaller \cite{greenlandFundamentalsEpidemiologicData2008}. Adding that approximation error to the usual study uncertainties, we do not expect \textit{P}-values below 0.001 from z-scores to have more than 2-digit accuracy, and thus advise rounding \textit{S}-values above $−\log_{2}(0.001)$ $\approx$ 10 to the nearest integer.

The \textit{S}-values for testing the same hypothesis from \textit{K} independent studies can be summed to provide a summary test statistic for the hypothesis (see \href{https://arxiv.org/abs/2008.12991}{Supplement} \cite{greenlandTechnicalIssuesInterpretation2020}). A caution is needed in that the resulting sum will have an expectation equal to \textit{K} under the hypothesis and background assumptions. Thus its size must be evaluated against a distribution that increases with \textit{K} (specifically, by doubling the sum and comparing it to a $\chi^{2}$ distribution with 2\textit{K} degrees of freedom) \cite{coxChapterPureSignificance1974,greenlandValidPvaluesBehave2019}. 

As discussed in the \href{https://arxiv.org/abs/2008.12991}{Supplement} \cite{greenlandTechnicalIssuesInterpretation2020}, in Bayesian settings one may see certain \textit{P}-values that are not valid frequentist \textit{P}-values, the primary example being the posterior predictive \textit{P}-value \cite{bayarriValuesCompositeNull2000,robinsAsymptoticDistributionValues2000}; unfortunately, the negative logs of such invalid \textit{P}-values do not measure surprisal at the statistic given the model, and so are not valid \textit{S}-values. 

As mentioned earlier, one purpose of converting \textit{P}-values to \textit{S}-values is to thwart the fallacy of mistaking data probabilities like a \textit{P}-value for hypotheses probabilities. It is often said that this fallacy is addressed by Bayesian methods because they give the reader hypothesis probabilities. A problem with such probabilities is that deriving them requires the analyst to supply a prior distribution (“prior”) that supplies initial probabilities for competing hypotheses. In many serious applications, there is no simple, universal, and reliable guide to choosing a prior (other than as a shrinkage/penalization/regularization device to improve certain frequency properties), and thus posterior probability statements can vary considerably across analysts even when there is no disagreement about frequentist results \cite{starkConstraintsPriors2015}). That problem is precisely why frequentists reject Bayesian methods as a general foundation for data analysis.

In sharp contrast, frequency models for the data can be enforced by experimental devices, producing information that can be quantified even without agreement about a prior distribution for targeted quantities. This quantification does not preclude a further analysis which combines the experimental information with external information encoded in a penalty function or prior distribution (which may be partial \cite{coxNotePartiallyBayes1975}). Nor does it free data analysts from responsibility to weaken their interpretations when using methods derived from devices or assumptions that are not known to be operative \cite{greenlandAidScientificInference2020}. For example, explanations for results from randomization tests in nonrandomized studies must include not only treatment effects and random error among possible explanations, but also effects of randomization failure \cite{greenlandRandomizationStatisticsCausal1990,greenlandIdentifiabilityExchangeabilityEpidemiological1986}. Finally, we caution that Gelman and Carlin \cite{Gelman2014-qk} refer to erroneously inferring the wrong sign of a parameter as “type-S error”, an entirely different usage of “S”. 
\vspace{-.80cm}
\noindent\textcolor{white}{\rule[-0.5ex]{\linewidth}{0.05pt}}
\vspace{-.50cm}
\subsection[Tests of different values for a parameter vs. tests of different parameters]{\texorpdfstring{Tests of different values for a parameter vs. tests of different parameters}{Tests of different values for a parameter vs. tests of different parameters}}

Even if all background assumptions hold, no single number (whether a \textit{P}-value, \textit{S}-value, or point estimate) can by itself provide an adequate measure of sample information about a targeted parameter, such as a mean difference, a hazard ratio (HR), or some other contrast across treatment groups. We have thus formulated our description to allow the test hypothesis \textbf{H} to refer to different values for the same parameter. For example, \textbf{H} could be “HR = 1”, the traditional null hypothesis of no change in hazard rate across compared groups; but \textbf{H} could just as well be “HR = 2”, or “HR $\leq$ 2”, or even “$\nicefrac{1}{2}$  $\leq$ HR $\leq$ 2” \cite{greenlandValidPvaluesBehave2019}. In all these variations, the set of auxiliary assumptions (background model) used to compute the statistics stays unchanged; only \textbf{H} is changing. Unconditionally, the \textit{S}-values for the different \textbf{H} are measuring information against different restrictions on HR beyond the background assumptions, which stay the same. \\ \indent
A similar comment applies when, in a model, we test different coefficients: The background assumptions are unchanged, only the targeted test hypothesis \textbf{H} is changing, although now the change is to \textit{another parameter} (rather than another value for the same parameter). For example, in a model for effects of cancer treatments we might compute the \textit{P}-value and \textit{S}-value from a test of $\textbf{H}_{r}$ = “the coefficient of radiotherapy is zero” and another \textit{P}-value and \textit{S}-value from a test of $\textbf{H}_{c}$ = “the coefficient of chemotherapy is zero.” Conditionally, these two \textit{S}-values are giving information against different target hypotheses $\textbf{H}_{r}$ and $\textbf{H}_{c}$ using the same background model; for example, using a proportional-hazards model, that background includes the assumption that the effects of different treatments on the hazard multiply together to produce the total effect of all treatments combined. Unconditionally, these \textit{S}-values are measuring information against different test models: a model with no effect of radiotherapy but allowing an effect of chemotherapy, and a model allowing an effect of radiotherapy but no effect of chemotherapy; all other assumptions are the same in both models (including possibly unseen and inappropriate assumptions about causal ordering \cite{westreichTableFallacyPresenting2013}). \\ \indent
Testing different parameters with the same data raises issues of multiple comparisons (also known as simultaneous inference). These issues are very complex and controversial, with opinions about multiple-comparison adjustment ranging from complete dismissal of adjustments to demands for mindless, routine use, and extend far beyond the present scope; see \cite{greenlandMultipleComparisonsControversies2019,greenlandAnalysisGoalsErrorcost2020} for a recent commentary and review. We can only note here that the devices we recommend can also be applied to adjusted comparisons; for example, the \textit{S}-value computed from an adjusted \textit{P}-value becomes the information against a hypothesis penalized (reduced) to account for multiplicity. \\ \indent
We caution however against confusing the problem of testing multiple parameters with the testing of multiple values of the \textit{same} parameter, as we recommend here: Tests of the same parameter are logically dependent in a manner eliminating the need for adjustment. This dependency can be seen in how a \textit{P}-value for HR $\leq$ 1 must be less than the \textit{P}-value for the less restrictive HR $\leq$ 2 (using a test derived from the same method and assumptions). Note also that a compatibility interval requires selection of values based on multiple tests of the parameter, namely the values for which $p>\alpha$; this selection does not harm any frequency property of the interval (e.g., coverage of the true parameter value at a rate 1 − $\alpha$ if all background assumptions are correct). 
\vspace{-.50cm}
\noindent\textcolor{white}{\rule[-0.5ex]{\linewidth}{0.05pt}}
\vspace{-.90cm}
\section{Conclusion}

\vspace{-.40cm}
Ongoing misinterpretations of important medical research demonstrate the need for simple reforms to traditional terms and interpretations. As lamented elsewhere, \cite{amrheinScientistsRiseStatistical2019,amrheinInferentialStatisticsDescriptive2019,mcshaneStatisticalSignificanceDichotomization2017,mcshaneAbandonStatisticalSignificance2019}, those traditions have led to overinterpretations and misinterpretations becoming standards of reporting in leading medical journals, with ardent defense of such malpractice by those invested in the traditions. Especially when there is doubt about conventional assumptions, overconfident terms like “significance,” “confidence,” and “severity” and decisive interpretations should be replaced with more cautiously graded unconditional descriptions such as “compatibility”; narrowly compressed probabilities like \textit{P}-values can be supplemented with quantitative-information concepts like \textit{S}-values; and requests can be made for tables or graphs of \textit{P}-values and \textit{S}-values for multiple alternative hypotheses, rather than forcing focus onto null hypotheses \cite{folksIdeasStatistics1981,greenlandValidPvaluesBehave2019,Poole1987-nb,Sullivan1990-ha}. These reforms need to be given a serious chance via editorial encouragement in both review and instructions to authors. \\
\vspace{-.15cm}
\noindent\textcolor{white}{\rule[-0.5ex]{\linewidth}{0.05pt}}
 
\small
\vspace{-.40cm}
\noindent\textcolor{gray!50}{\rule[-0.5ex]{\linewidth}{0.05pt}}
\vspace{-.80cm}
\subsection*{Acknowledgements}
\vspace{-.15cm}
We are most grateful for the generous comments and criticisms on our previous drafts offered by Andrew Althouse, Valentin Amrhein, Andrew Brown, Darren Dahly, Frank Harrell Jr, John Ioannidis, Daniël Lakens, Nicole Lazar, Gregory Lopez, Oliver Maclaren, Blake McShane, Tim Morris, Keith O’Rourke, Kristin Sainani, Allen Schirm, Philip Stark, Andrew Vickers, Andrew Vigotsky, Jack Wilkinson, Corey Yanofsky, and the referees. We also thank Karen Pendergrass for her help in producing the figures in this paper. Our acknowledgment does \textit{\textbf{not}} imply endorsement of our views by these colleagues, and we remain \textit{solely responsible} for the views expressed herein.
\vspace{-.5cm}
\subsection*{Authors' Contributions}
\vspace{-.15cm}
Both authors ($\textbf{ZR}^{1}$ and $\textbf{SG}^{2}$) wrote the first draft and revised the manuscript, read and approved the submitted manuscript, and have agreed to be personally accountable for their own contributions related to the accuracy and integrity of any part of the work.

\subsection*{Abbreviations}

\textbf{A}: Background model assumptions; \textbf{ASD}: Autism spectrum disorder; \textbf{CI}: Compatibility/confidence interval; \textbf{H}: Test hypothesis; \textbf{HDPS}: High-dimensional propensity score; \textbf{HR}: Hazard ratio; \textbf{LI}: Likelihood interval; \textbf{LR}: Likelihood ratio; \textbf{M}: Test model; \textbf{MLR}: Maximum-likelihood ratio; \textbf{NHST}: Null-hypothesis significance test; \textbf{\textit{S}-value}: Surprisal (Shannon-information) value

\noindent\textcolor{gray!50}{\rule[-0.5ex]{\linewidth}{0.05pt}}
\vspace{-.70cm}
\subsection*{Data and Materials}
\label{sec:Data}

\begin{flushleft}
The datasets used and analyzed in the current paper are available in the \href{https://osf.io/6w8g9/}{Open Science Framework} |\\ DOI: \href{https://osf.io/6w8g9/}{10.17605/OSF.IO/6W8G9} 
\end{flushleft}
\vspace{-.50cm}
\noindent\textcolor{gray!50}{\rule[-0.5ex]{\linewidth}{0.05pt}}
\vspace{-.80cm}
\subsection*{Article Citation}
\otherinfo{Rafi, Z., Greenland, S. Semantic and cognitive tools to aid statistical science: replace confidence and significance by compatibility and surprise. \\\textit{BMC Med Res Methodol} \textbf{20}, 244 (2020). \\ \href{https://doi.org/10.1186/s12874-020-01105-9}{https://doi.org/10.1186/s12874-020-01105-9}}
\vspace{-.40cm}
\noindent\textcolor{gray!50}{\rule[-0.5ex]{\linewidth}{0.05pt}}
\vspace{-.80cm}
{\small
\bibliographystyle{naturemag}
\bibliography{References.bib}
}

\end{multicols}

\newpage

\begin{adjustwidth*}{60pt}{60pt}
\noindent\textcolor{gray!50}{\rule[-0.5ex]{\linewidth}{0.05pt}}
\section{Appendix}
\label{sec:Appendix}

\noindent\textcolor{gray!50}{\rule[-0.5ex]{\linewidth}{0.05pt}}

\subsection[Technical details of computations for the tables]{\texorpdfstring{Technical details of computations \\for the tables}{Technical details of computations for the tables}}
\vspace{-.250cm}
\noindent\textcolor{gray!50}{\rule[-0.5ex]{\linewidth}{0.05pt}} \\

\vspace{-.10cm}

{\fontsize{9}{20}\selectfont{
\noindent
\textbf{Table~\ref{Tab:tab1}} shows relations under the standard 1 degree-of-freedom (df) $\chi^{2}$ approximation for the LR statistic when \textbf{H} is a hypothesis that a parameter equals a specific value, e.g., for the hypothesis that a hazard ratio HR equals the number r, \textbf{H}: HR = r. For normal (Gaussian) data these relations are exact and the LR statistic reduces to squared z-score for the hypothesis \cite{cummingsAnalysisIncidenceRates2019}. The \textit{S}-value and LR statistic track each other rather closely although the latter increases more rapidly. Their relation reflects that, under the test model and the standard approximations, the \textit{P}-value is uniform and hence the \textit{S}-value is unit-exponential, which is half a 2 df $\chi^{2}$ \cite{fisherStatisticalMethodsResearch1925} and hence has a heavier right tail than the 1 df LR statistic; specifically, with $x=\ln(r)$, the ratio of densities for the 2 df and 1 df $\chi^{2}$ is proportional to $x^{\nicefrac{1}{2}}$.}} \\ 

\vspace{-.50cm}
\noindent\textcolor{gray!50}{\rule[-0.5ex]{\linewidth}{0.05pt}} \\

\vspace{-.10cm}

{\fontsize{9}{20}\selectfont{
\noindent
For \textbf{Table~\ref{Tab:tab2}} and the figures, statistics were computed from the approximate normal distribution used for the CIs in Brown et al. \cite{brownAssociationSerotonergicAntidepressant2017}, in which the log-hazard ratio $\ln(HR)$ is estimated to have mean $m=\ln(1.61)$ and standard deviation $d=\frac{\ln(\frac{2.59}{.997})}{2(1.96)}$. The \textit{P}-value for \textbf{H}: HR = r is then derived from the normal score $Z=\frac{\ln(\frac{1.61}{r})}{d}$, and the LR statistic and MLR are approximated by $Z^{2}$ and $\exp(\frac{Z^{2}}{2})$. For contrast to the \textit{P}-graph in \textbf{Figure \ref{Fig2}}, \textbf{Figure S\ref{SuppFig1}} shows the relative likelihood function, $\nicefrac{1}{MLR}$, produced from the Brown et al. HDPS results, taking the maximum as the reference point so that the graph extends from 0 to 1. It may be noticed that this function appears proportional to a posterior probability density for $\ln(HR)$, but this proportionality holds only under very special conditions. For contrast to the \textit{S}-graph in \textbf{Figure \ref{Fig3}}, \textbf{Figure S\ref{SuppFig2}} shows the corresponding deviance function $2\ln(MLR)$. }} \\
\vspace{.20cm}
\noindent\textcolor{white}{\rule[-0.5ex]{\linewidth}{0.05pt}}

\newpage

\begin{suppfigure*}[t]
\section{Supplementary Figures}
\vspace{-.75cm}
\noindent\textcolor{white}{\rule[-0.5ex]{\linewidth}{0.05pt}}
\vspace{-.35cm}
\subsection{Relative Likelihood Function}
\begin{center}
\fbox{\includegraphics[width=25pc, height=15pc]{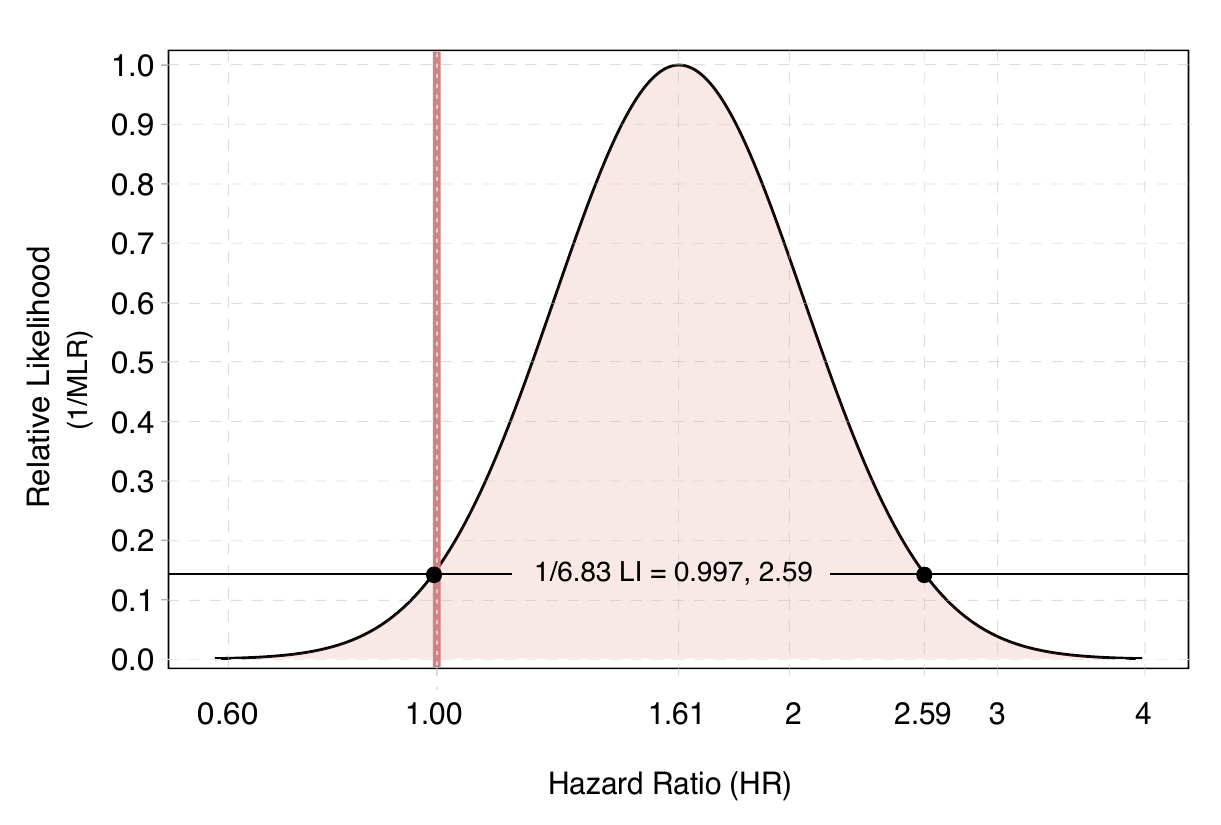}}
\captionsetup{width=.75\linewidth}
\caption{\small\textbf{Relative likelihoods for a range of hazard ratios.} A relative likelihood function that corresponds to \textbf{Figure \ref{Fig2}}, the \textit{P}-value function. Also plotted is the $\nicefrac{1}{6.83}$ likelihood interval (\textbf{LI}), which corresponds to the 95\% compatibility interval. Computed from results in Brown et al. \cite{brownAssociationSerotonergicAntidepressant2017}.  \textbf{MLR} = Maximum-Likelihood Ratio. \textbf{HR} = 1 represents no association.}
\label{SuppFig1}
\begin{adjustwidth*}{60pt}{60pt}
\vspace{-.50cm}
\end{adjustwidth*}
\end{center}
\vspace{-.50cm}
\end{suppfigure*}

\begin{suppfigure*}[t]
\vspace{-.30cm}
\subsection{Deviance Function}
\begin{center}
\fbox{\includegraphics[width=25pc, height=15pc]{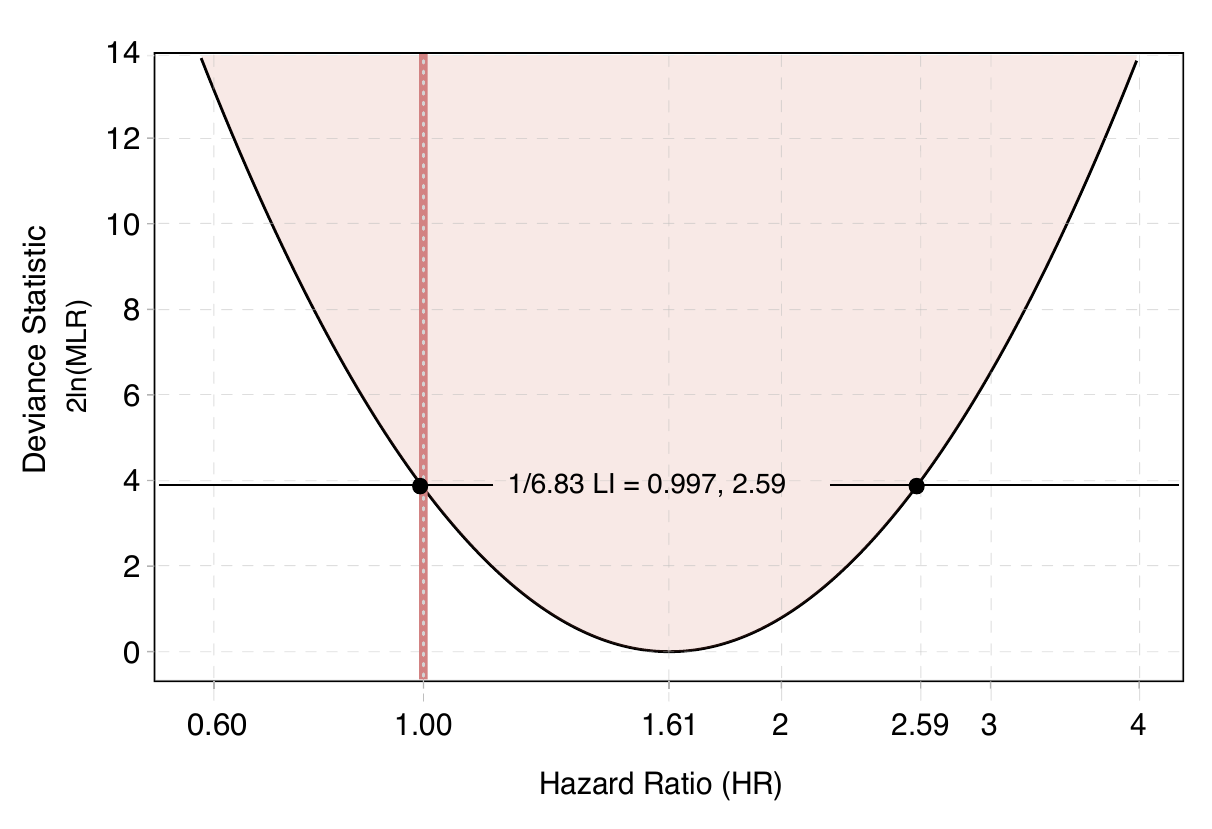}}
\captionsetup{width=.75\linewidth}
\caption{\small\textbf{Deviance statistics for a range of hazard ratios.} A deviance function, which corresponds to \textbf{Figure \ref{Fig3}}, the \textit{S}-value function. Also plotted is the likelihood interval (\textbf{LI}), which corresponds to the 95\% compatibility interval. Computed from results in Brown et al. \cite{brownAssociationSerotonergicAntidepressant2017}. \\\textbf{MLR} = Maximum-Likelihood Ratio. \textbf{HR} = 1 represents no association. }
\label{SuppFig2}
\end{center}
\end{suppfigure*}

\end{adjustwidth*}
\end{document}